\documentstyle[12pt,epsfig]{article}

\newcommand{\uf}{``up-flat" }   
\newcommand{\df}{``down-flat" }   
\newcommand{\us}{``up-steep" }   
\newcommand{\ds}{``down-steep" }   
\newcommand{\be}{\begin{equation}}   
\newcommand{\ee}{\end{equation}}   
\newcommand{\ba}{\begin{eqnarray}}   
\newcommand{\ea}{\end{eqnarray}}   
\newcommand{\mpp}{$m_{\pi\pi}$ }   
\newcommand{\pipi}{$\pi\pi$ }   
   
\newcommand{\pppm}{$\pi^{+}\pi^{-}$ }   
   
\newcommand{\popo}{$\pi^{0}\pi^{0}$ }

\title{\bf Roy's equations and the $\pi\pi$ experimental data}

\author{R. Kami\'nski$^a$, L. Le\'sniak$^a$ and B. Loiseau$^b$ \\
{\normalsize\it $^a$ Department of Theoretical Physics,}\\{\normalsize\it  H. Niewodnicza\'nski
Institute of Nuclear Physics,}\\ {\normalsize\it Polish Academy of Sciences, PL 31-342 Krak\'ow, Poland}\\
{\normalsize\it $^b$ Laboratoire de Physique Nucl\'eaire et de Hautes 
\'Energies}\footnote{Unit\'e de Recherche des Universit\'es
Paris 6 et Paris 7, associ\'ee au CNRS}, {\normalsize\it Groupe} \\{\normalsize\it Th\'eorie,
Univ. P. \& M. Curie, 4 Pl. Jussieu, F-75252 Paris, France}\footnote{This work has been performed in the
framework of the IN2P3-Polish Laboratories Convention (project number 99-97).} }

\begin{document}

\maketitle

\vspace{-1cm}

\begin{abstract}

Roy's equations are used to check if the scalar-isoscalar \pipi scattering 
amplitudes fitted to experimental data fulfill crossing symmetry conditions.
It is shown that the amplitudes describing the ``down-flat'' phase shift solution 
satisfy crossing symmetry below 1 GeV while the amplitudes fitted to the ``up-flat'' 
data do not. 
In this way the long standing "up-down" ambiguity in the
 phenomenological determination of the scalar-isoscalar
 \pipi amplitudes has been  resolved confirming the
independent result of  the recent joint analysis of the \pppm and \popo data.

\end{abstract}

\section{Introduction}

\hspace{0.6cm}In 1997 a new analysis of the $\pi^-p_\uparrow\to\pi^+\pi^-n$ reaction on a 
polarized target was performed in the \mpp effective mass range from 600  to 1600~MeV \cite{klr1}. 
For the first time the pseudoscalar ($\pi$-exchange) amplitude was separated from the pseudovector ($a_1$-exchange) 
amplitude in the region of the the four-momentum transfer squared from $-0.005$ to $-0.2$ (GeV/c)$^2$.
Below 1000~MeV, where the $S$- and $P$-waves strongly interfere, the partial wave analysis of the \pppm data provided  us
with two scalar-isoscalar solutions, called "up" and "down", which differ by their intensities.
Lack of information on a sign difference between the phases of the $S$- and $P$-waves near the position of the $\rho$
resonance led us to other two branches of the "up" and "down" amplitudes named "steep" and "flat".
It was shown in \cite{klr2} that both \us and \ds $S$-wave isoscalar amplitudes significantly violate unitarity 
below 1~GeV and should be rejected as nonphysical.
Two remaining "flat" amplitudes survived the unitarity check and other tests were needed to resolve 
the existing "up-down" ambiguity.

In 2001 new experimental data on the \popo production from the E852 collaboration appeared \cite{gunter01} and were
used in a joint analysis of the \pppm and \popo data \cite{klr3}.
The \popo data are very useful to compare with the \pppm data due to an absence of the $P$-wave in the \popo channel 
and therefore a lack of the "up-down" ambiguity.
The one-pion and $a_1$-exchange model described in \cite{klr3} was used to calculate the $S$-wave intensities of 
the \popo production by choosing as an input the \uf or \df phase shifts.
The isospin relations between the $\pi^+\pi^-\to\pi^+\pi^-$ and $\pi^+\pi^-\to\pi^0\pi^0$ amplitudes supplemented by the
parameterization of the isotensor scalar amplitude taken from \cite{klr1} were helpful in these calculations.
It was shown that the \popo $S$-wave intensities determined for the \df phase shifts agree with the experimental values
within the errors. However, for the \uf phase shifts in the \mpp range from 850 to 970~MeV important differences 
between the calculated \popo intensities and the corresponding experimental values occur.
This fact led the authors to a conclusion that the \uf data set should also be rejected.

\section{Roy's equations as a test for the \pipi amplitudes}

\hspace{0.6cm}Another independent test of the \uf and \df amplitudes consists in checking if they fulfill crossing symmetry conditions below 1~GeV. 
In order to achieve this task we have used Roy's equations \cite{roy71} for the scalar-isoscalar, $\ell=0,\ I=0$,
scalar-isotensor, $\ell=0,\ I=2$, and the vector-isovector, $\ell=1,\ I=1$,
$\pi\pi$ partial waves determined in a wide \mpp range.
We were especially interested in the \mpp region between 800 and 1000~MeV where differences between phase shifts 
of the \uf and \df data sets are largest and reach about 45$^{\rm o}$ (see Fig. 4 in \cite{klr2}).
In a recent analysis of Roy's equations \cite{anan}
a special attention was put on the effective mass lower than 800~MeV.

As an input in Roy's equations  we have used imaginary parts of 
the partial waves  amplitudes  $f_{\ell}^{I}(s)$  related to the 
$\pi \pi$ phase shifts $\delta_{\ell}^{I}$ and inelasticities $\eta_{\ell}^{I}$:

\begin{equation}
    f_{\ell}^{I}(s)=\sqrt{\frac{s}{s-4\mu^2}}\frac{1}{2i}
    \left(\eta_{\ell}^{I}e^{2i\delta_{\ell}^{I}}-1\right),
    \label{eq:3}
\end{equation}
where $\mu$ is the charged pion mass and $s = m^2_{\pi\pi}$.

Below 970~MeV the following Pad\'e representation of both the ``up-flat" and ``down-flat" phase shifts 
has been taken:
\begin{equation}
    \tan\delta_{0}^0(s) = \frac{\sum_{i=0}^{4}\alpha_{2i+1}k^{2i+1}}
     {\Pi_{i=1}^{3}(k^2/\alpha_{2i}-1)},
    \label{eq:Pade}
\end{equation}
where $k=\frac{1}{2}\sqrt{s-4\mu^2}$ is the  pion  momentum and 
$\alpha_{j}\,(j=1,\ldots,7,9)$ are constant parameters. 
Above 970~MeV up to 2 GeV our coupled channel model  \cite{kamplb97} amplitude A, fitted to the \df
data, and 
the amplitude C, constrained by the \uf data, were applied.
In the fits we also used the near threshold phase shifts calculated from the differences $\delta_0^0-\delta_1^1$
obtained in the high statistics $K_{e4}$ decay experiment \cite{pislak01}. 
The scattering length $a_0^0$ and the slope parameter $b_0^0$ are directly related to the constants $\alpha_j$:
$a_{0}^0 = -\alpha_{1}\mu $ and 
$b_{0}^0 = -\alpha_{1}\mu \left ( 0.5\mu^{-2}+\alpha_{2}^{-1}+\alpha_{4}^{-1}+\alpha_{6}^{-1}
-\alpha_{1}^2\right) - \alpha_{3}\mu$.

The parameterization of isotensor wave using a rank-two separable potential model  has been described 
in \cite{kll03} where detailed analysis of the present study is presented.

For the $P$-wave, from the \pipi threshold till 970~MeV, we have used an extended Schenk parameterization 
\cite{anan}:
\begin{equation}
\tan \,\delta_1^1(s) = 
\sqrt{1-\frac{4\mu^2}{s}}
\ k^2\left(A+Bk^2+Ck^4+Dk^6\right)\left(\frac{4\mu^2-s_\rho}{s-s_\rho}\right),
\label{eq:schenk}
\end{equation}
where $A$ is the $P$-wave scattering length and $s_\rho$ is equal to the $\rho$-mass squared. 
Above 970~MeV we took the $K$-matrix parameterization of Hyams et al. \cite{hyams73}. 
The parameters $C$ and $D$ were chosen to join smoothly the phase shifts given by both parameterizations around 970~MeV.

The contributions to Roy's equations from high energies $(m_{\pi\pi} > 2$ GeV) and from higher partial waves 
($l>1$) are called driving terms.
They are composed of contributions from the $f_{2}(1270)$ and $\rho_{3}(1690)$ resonances and 
from the Regge amplitudes for the Pomeron, $\rho$- and $f$-exchanges.
The Breit-Wigner parameterization with masses, widths and $\pi\pi$ branching ratios taken from \cite{pdg02}
were used for $f_{2}(1270)$ and $\rho_{3}(1690)$.
For the Regge parts we have used formulae of \cite{anan} without the $u$-crossed terms.
We have found that the $f_{2}(1270)$ resonance dominates in the scalar isoscalar wave and that the introduction of the
$\rho_{3}(1690)$ has a significant influence on the isotensor and isovector waves.
In the isoscalar wave the Regge contributions are more than 10 times smaller than the resonance contributions
but for the isospin 1 and 2 they are of the same order.

The thirteen constants, six for the scalar-isoscalar wave in (\ref{eq:Pade}), four for the isotensor
wave and three for the isovector wave in (\ref{eq:schenk}), 
were calculated from the simultaneous fits to data and to Roy's equations  separately for the \uf and \df data. 
We have used the CERN MINUIT program with the $\chi^2$ test function defined by 
\begin{equation}
\chi^2 = \sum_{I=0,1,2} \left\lbrace\sum_{i=1}^{N_I} 
\left[ \frac{\sin \left(\delta_{\ell}^I \left(s_i\right) 
- \varphi_{\ell}^I \left(s_i\right) \right)}
       {\Delta \varphi_{\ell}^I \left(s_i\right)} \right]^2 +
\sum_{j=1}^{12} 
	\left [ \frac
   {\mbox{Re }f_{out}^I\left(s_{j}\right)-\mbox{Re }f_{in}^I\left(s_{j}\right)}
	{\Delta f} \right]^2 \right\rbrace,
        \label{eq:chi2}
\end{equation}
where $\varphi_{\ell}^I \left(s_i\right)$ and 
$\Delta \varphi_{\ell}^I \left(s_i\right)$ represent the experimental phase 
shifts and their errors, respectively, 
$s_j=[4j+0.001]\mu^2$ for $j=1, ..., 11$ and $s_{12}=46.001\mu^2$.
The real parts Re~$f_{in}^I$ have been calculated from (\ref{eq:3}) under an assumption that the inelasticity $\eta^I_l$ is
equal to 1 and the phase shifts $\delta^I_l$ are equal to $\phi^I_l(s_j)$.
Other real parts, denoted by Re~$f^I_{out}$, constitute the output values calculated from Roy's equations.
We take a $\Delta f$ value of $0.5\times 10^{-2}$ to obtain acceptable fits to Roy's equations. 
18 experimental values of the \uf or \df data between 600 and 950~MeV were used in addition to six 
data taken from
\cite{pislak01}.

In Fig. 1a and 1b we present results of fits to the \uf and \df phase shifts and to Roy's equations (solid
lines). 
In both cases differences $\mid$~Re~$f^I_{out}$~-~Re~$f^I_{in}$~$\mid$
were of the order of $10^{-3}$ in all three partial waves.
The $\chi^2$ for 18 points between 600 and 970~MeV was 16.6 in the \df case and as large as 46.4 in the \uf one.
We see in Fig. 1a that the solid line lies distinctly below the \uf data points between 
800 and 970~MeV. 
In contrary, the corresponding line for the \df case in Fig. 1b is very close to experimental data in the same range of \mpp.
In order to improve a fit to the \uf data we have used constraints given by the good fit to the \df data.
Two parameters were fixed by choosing the values of the scattering length and the slope parameter and two others by the values of phase 
shifts calculated from this fit at 500 and 550~MeV. 
A new fit with these constrains gave an improved value of $\chi^2=13$ for 18 \uf data points, corresponding to the first
part of $\chi^2$ in (\ref{eq:chi2}) but provided us with an enormous value of
$\chi^2 = 1.2\times 10^4$ for the second part related to Roy's equations.
The phase shifts for this amplitude are presented in Fig. 1a by the dotted line.
It is clear that a simultaneous good fit to the \uf data and to Roy's equations is impossible.

\begin{figure}[h!]

\vspace{2cm}

\hspace{-1cm}\mbox{\epsfxsize 16.5cm\epsfysize 14.cm\epsfbox{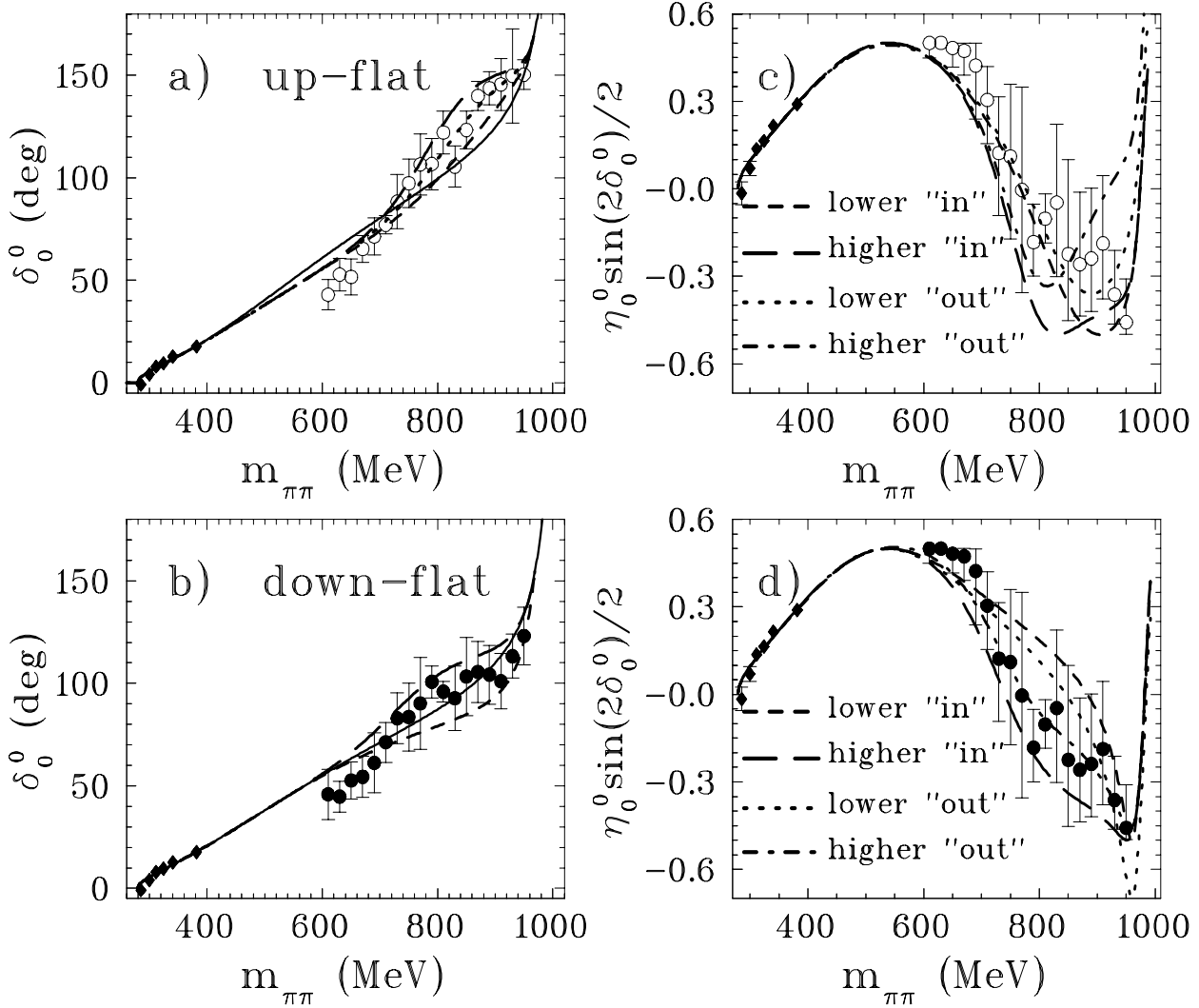}}\\
\caption{\textbf{a)} and \textbf{c)} correspond to the \uf case, \textbf{b)} and \textbf{d)} correspond to the \df case.
Fits to the scalar-isoscalar phase shifts of \cite{klr1} and to Roy's equations are denoted by
solid lines in \textbf{a)} and \textbf{b)}.
Dotted line in {\bf a)} between two dashed lines represents fit with constraints described in the text.
Dashed lines in \textbf{a)} and \textbf{b)} represent fits to phase shifts moved upwards and downwards by their errors;
the corresponding lines in  \textbf{c)} and \textbf{d)} are called {\it higher} and {\it lower}, respectively.
Lines in \textbf{c)} and \textbf{d)} correspond to real parts of input amplitudes {\it ("in")} and real parts
calculated from 
Roy's equations {\it ("out")}, all multiplied by $2ks^{-1/2}$.
Diamonds represent the K$_{e4}$ data~\cite{pislak01}.
}
\end{figure}


Apart of the fits to the \uf and \df experimental points we~have~also~performed fits to points shifted upwards
and downwards by their errors.
In these fits the same four constraints described above were used below 600~MeV.
Up to 937~MeV in the \df case in Fig. 1d the curves labeled {\it higher "in"} and {\it lower "in"} form a band including inside a band delimited
by the lines {\it higher "out"} and {\it lower "out"}.
All the curves lying inside these bands correspond to the amplitudes fulfilling the crossing symmetry so the \df data can
be accepted as physical ones.
In the \uf case in Fig. 1c the output band lies outside of the input band from 840 to 970~MeV.
It means that in this case the crossing symmetry is violated by the amplitudes fitted to the \uf data.

In Fig. 2 we have presented the output results  for the isotensor and isovector waves in the \df case only since
in the \uf case the curves are very similar.
The "in" curves were not plotted because they are almost indistinguishable from the "out" ones.

\begin{figure}
\mbox{\epsfxsize 15.5cm\epsfysize 7.cm\epsfbox{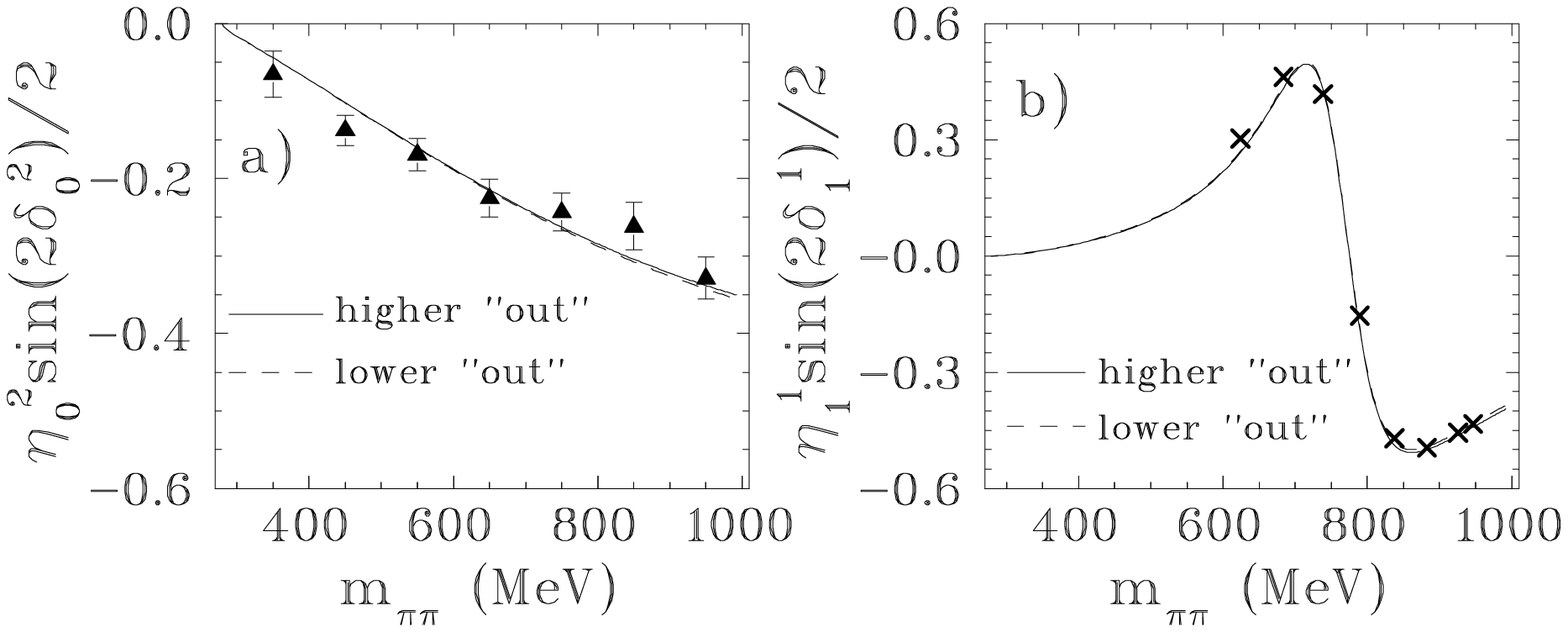}}\\
\caption{Real parts of isotensor {\bf a)} and isovector {\bf b)} \pipi amplitudes (multiplied by $2ks^{-1/2}$) 
fitted to the \df data. Triangles in {\bf a)} denote data of~\cite{hoogland77}.
Crosses in  {\bf b)} are the pseudo-data calculated from the $K$-matrix fit of~\cite{hyams73}.}
\end{figure}

\section{Conclusions}

\hspace{0.6cm}We have used Roy's equations as a tool to test if the amplitudes fitted to the \uf and \df phase shifts 
extracted from the $\pi^-p_\uparrow\to\pi^+\pi^-n$  data fulfill crossing symmetry conditions. 
We have found that only the $S$-wave isoscalar amplitude corresponding to the \df data set can be accepted.
The amplitude constrained to the \uf data does not satisfy Roy's equations  and should be rejected as nonphysical.
This conclusion is in agreement with the independent results obtained from a joint analysis of the $\pi^+\pi^-$ and 
the $\pi^0\pi^0$ production data \cite{klr3}.
In this way a long standing "up-down" ambiguity in the \pipi experimental data has been resolved in favour of 
the \df data set.


\begin{thebibliography}{99}

    \bibitem{klr1}  Kami\'nski R., Le\'sniak L. and Rybicki K., Z. Phys. 
     {\bf C 74}, 79 (1997).

    \bibitem{klr2}   Kami\'nski R., Le\'sniak L. and Rybicki K., Acta Phys. 
    Pol. \textbf{B31}, 895 (2000).

    \bibitem{gunter01}  Gunter J. et al. (E852 Collaboration), Phys. 
    Rev. {\bf D 64},  072003 (2001).

    \bibitem{klr3} Kami\'nski R., Le\'sniak L. and Rybicki K.,
    Eur. Phys. J. direct \textbf{C4}, 1 (2002).

    \bibitem{roy71}  Roy S. M., Phys. Lett. {\bf B 36}, 353 (1971);
    Roy S. M.,  Helv. Phys. Acta {\bf 63}, 627  (1990).

    \bibitem{anan}  Ananthanarayan B., Colangelo G., Gasser J. and 
    Leutwyler H., Phys. Rep. {\bf 353}, 207 (2001).

    \bibitem{kamplb97} Kami\'nski R., Le\'sniak L., B. Loiseau, Phys. 
    Lett. {\bf B 413}, 13 (1997).

     \bibitem{pislak01} Pislak S. {\it et al.}, (E865 Coll.),
     Phys. Rev. Lett. {\bf 87}, 221801 (2001).

    \bibitem{kll03} Kami\'nski R., Le\'sniak L. and B. Loiseau, Phys. Lett. 
     {\bf B 551}, 241 (2003).

    \bibitem{hyams73} Hyams B. {\it et al.}, Nucl. Phys. {\bf B 64}, 134 (1973).
   
    \bibitem{pdg02} K. Hagiwara {\it et al.} (Particle Data Group), 
    Phys. Rev. \textbf{D66}, 010001 (2002).

    \bibitem{hoogland77} Hoogland W. {\it et al.},
      Nucl. Phys. \textbf{B126}, 109  (1977).


\end{thebibliography}
\end{document}